\documentclass[runningheads]{llncs}
\usepackage{fullpage}
\usepackage{graphicx}
\usepackage{booktabs}
\usepackage{tabularx}
\usepackage{lipsum}  %% Package to create dummy text (comment or erase before start)
\usepackage{color}
\usepackage{etoolbox}
\usepackage{appendix}
\usepackage{hyperref}
\usepackage{bbm}
\usepackage{enumitem}
\usepackage{amsmath}
\usepackage{amssymb}
\usepackage{comment}

\begin{document}
\title{Choice-Aware User Engagement Modeling and Optimization on Social Media}

\author{Saketh Reddy Karra\inst{1}
\and
Theja Tulabandhula\inst{2}
} 
\authorrunning{Saketh Reddy Karra\inst{1} \and  Theja Tulabandhula\inst{2}}

\institute{University of Illinois at Chicago, Chicago IL 60607, USA\\ \email{skarra7@uic.edu}\and
University of Illinois at Chicago, Chicago IL 60607, USA \\
\email{theja@uic.edu}}
\maketitle              
\begin{abstract}
We address the problem of maximizing user engagement with content (in the form of like, reply, retweet, and retweet with comments) on the Twitter platform. We formulate the engagement forecasting task as a multi-label classification problem that captures choice behavior on an unsupervised clustering of tweet-topics. We propose a neural network architecture that incorporates user engagement history and predicts choice conditional on this context. We study the impact of recommending tweets on engagement outcomes by solving an appropriately defined tweet optimization problem based on the proposed model using a large dataset obtained from Twitter.

\end{abstract}
\section{Introduction}

%what is tweeting
Twitter is one of the most popular social networking sites, frequented by millions of users every day. The platform allows users to publish posts and share information publicly in the form of text messages, which are called \emph{tweets}.  For most users, the \emph{home timeline} view/screen is the default starting point, where a collection of tweets are displayed explicitly from accounts the users have chosen to follow. 

%what is the problem: user engagement
Users are often overwhelmed by the increasing number of tweets they receive daily, leading to information overload\cite{bontcheva2013social}. Many of the tweets that can be shown to the user are sub-optimal from an user engagement perspective, which is a set of metrics that focus on how the users interact with the content on the platform, how much time they spend, etc. Among all tweets by users the current user is following, due to the platform's constraint on how many tweets can be displayed on the home timeline, only a subset can actually influence the engagement metrics. In other words, even relevant posts could get sidelined and overlooked by users, leading to a long-term negative impact on engagement (for example, users might give up reading their timelines and start spending less time on the platform). Hence it is crucial to customize the home timeline with the most relevant tweets reflecting user's time-varying interests and enhance their experience on the platform.

% optimizing for the relevant tweets
On the home timeline view/screen, tweets can appear in an algorithmically ranked order or the reverse chronological order, depending on a pre-defined user preference. In the former case, all relevant tweets are scored by a predictive model that considers the user's interests and the relevance of the content. The most relevant tweets are displayed on their home timeline in the ranked order of scores. With the home timeline being an integral part of user experience on the platform, the quality of the predictive model that ranks these tweets directly reflects the quality of the user experience. Users tend to have more interactions and are more likely to return to the platform when their timelines are optimized to show the most relevant tweets first.

%optimizing is hard
Recommending the most engaging set of tweets is a combinatorial problem, and can be a difficult task to scale with many practical challenges. Naturally, users’ interest in a tweet or their willingness to read and react to a tweet is determined by many factors, such as the quality of the tweet, their personal interests, and the publisher’s authority among others. Further, trending topics change quite rapidly in response to events happening across the world. As a result, the underlying prediction models might become stale very quickly. While information from a user’s tweet history can and should be used to improve tweet recommendations, naively using all tweets created (e.g., since the platform’s launch or from the first sign-up times of the users) to model user behavior would be intractable and computationally very expensive. 

%feedback is hard
In general, users interact with content in different forms (e.g., these can be bucketed into categories such as \emph{Like, Retweet, Retweet with comment, and Reply}). This feedback from users on the displayed content is considered \emph{implicit} (as opposed to explicit feedback, such as the ratings and textual responses found on review websites). Users do not get to exclusively rate tweet relevance on their timeline due to various user experience constraints. Thus, the lack of explicit feedback makes recommending tweets more challenging. 

%what we do
Our work provides a novel way to forecast and optimize the subset of tweets that might be of interest to each user, given the user's overall timeline of interactions and other metadata. In particular, we propose a new approach based on a choice modeling framework to predict user engagement, and use it in an optimization scheme to recommend personalized tweets. In particular, we are able to model how a user engages with one tweet in the presence of other relevant tweets (thus modeling choice). This is a novel approach that complements many existing approaches and captures user behavior that is typically missed: users navigate the home timeline (e.g., scrolling up/down) and only interact with one or more tweets among multiple that are shown to them simultaneously. Thus we capture choice behavior because we can mode a tweet becoming more relevant or less relevant depending on other shown tweets.

We formulate the engagement forecasting task as a multi-label classification problem that captures choice behavior on an unsupervised clustering of \emph{tweet-topics}. This clustering is important to managing the scale of the unique tweets that appear on the platform. The corresponding neural network architecture also incorporates user engagement history in a partially interpretable manner. Given this choice-aware model of user behavior, we use a straightforward greedy approach (noting that the optimization problem is NP-hard) to optimize for the best set of personalized tweet-topics. These topics are then converted to tweets (by picking representative tweets via sampling), which are ultimately shown to the user. We report superior performance of the proposed choice-aware engagement model using a large scale dataset released by Twitter~\cite{belli2020privacyaware}.

The paper is structured as follows. In Section \ref{related work} we discuss closely related work, in Section \ref{topicchoicemodel} we introduce the  tweet-topic based choice model, and describe the proposed neural network architecture that predicts user engagement. In Section \ref{optimization}, we discuss the optimization setup for creating a personalized collection of tweets that maximize engagement using the aforementioned model. In Section \ref{experiments} we present the experiments to validate the proposed model, showing compelling performance results. We conclude in Section \ref{conclusion} with some comments on future work.

\section{Related work}
\label{related work}

A wide variety of recommendation methods have been proposed in the literature for micro-blogging sites such as Twitter. Over the past few years, among the multitude of work related to social media and recommendations systems, the following sample of research directions has focused on tasks that are similar to the one studied here: providing news feed and URL recommendations\cite{chen2010short,de2012chatter}, suggesting tweets and hashtags\cite{zangerle2011recommending}, suggesting users to follow\cite{hannon2010recommending}, and so forth. We briefly discuss the most relevant of these prior works next, and note in passing that there are many other works that look at similar problems in the Twitter context that we have to omit for brevity.

In \cite{duan2010empirical}, the authors develop a ranking based strategy that uses content relevance, account authority, and tweet-specific features to predict relevance scores for tweets. In \cite{uysal2011user}, the authors construct a tweet ranking model making use of re-tweeting behavior of users. The tweets and the users are both ranked based on their likelihood of getting a tweet re-tweeted. Our proposed model is different from these approaches as their ranking models are not contextualized with respect to other tweets and do not capture how they influence each other and user choice. Further, they also do not capture the dynamic change in the user’s interests.

In \cite{pennacchiotti2012making}, the authors design a model that leverages the content of the user’s tweets and those of his/her friends to recommend the most interesting tweets. In \cite{chen2012collaborative}, the authors propose a collaborative ranking model which considers tweet-topic level factors, user social relation factors, and explicit features such as authority of the publisher and quality of the tweet. Although both works use rich features, they do not explicitly account for user choice behavior.

Along similar lines, in \cite{naveed2011bad}, the authors discuss a method based on content features to predict the probability of a tweet being retweeted. The authors of \cite{elmongui2015trupi} built a personalized recommender system that combines users' social features and interactions, as well as their tweeting history. Further, they capture the time varying nature of user interests in different topics while generating engaging recommendations. Our proposed model predicts engagements without explicitly creating user interest profiles, which is common with most prior works. Instead, we use the idea of \emph{embeddings} (vector representations) to automatically encode behavior relevant to engagement. We also capture dynamic changes to such engagement behaviour by including recent engagement histories as explicit input features.

To summarise, in contrast to approaches discussed in the recommendation system literature and the above works, our work explicitly considers choice behavior, which has been extensively studied from both a statistical and computational standpoint in the marketing and revenue management (operations research) areas. We augment choice modeling with neural networks for our engagement maximization problem. In the marketing literature, \cite{gabel2020product} is a closely related work, where the authors propose a similar non-parametric model based on a custom neural network architecture to predict product choice over the entire assortment in a retail setting. Their solution is motivated by the coupon/discount optimization problem in retail. Given coupon assignments and customer purchase histories, their model predicts individual product choice probabilities for every product in a given assortment. Analogous to their work, we formulate an engagement maximization problem where showing a tweet is equivalent to displaying a coupon. Although the neural network  architecture that we propose is inspired by the product choice model in \cite{gabel2020product}, our treatment of tweets via tweet-topics to control complexity is completely new. We also work with a dataset that is much larger than the application considered in that paper, leading to some unique implementation challenges. Nonetheless, similar to their model, we define a comparative feed-forward neural network to model the tweet-topic engagement likelihoods.

\section{Choice-Aware Tweet Engagement Model}
\label{topicchoicemodel}

\subsection{Overview}

Given a large collection of tweets shown to ${I}$ users, we cluster them into $J$ topics using GSDMM \cite{yin2014dirichlet} topic modeling framework such that each cluster contains tweets corresponding to a singular topical theme. This process of mapping a tweet to a topical cluster allows us to work with a fixed/slowly growing number of topics instead of the underlying tweets themselves, as the latter can grow exponentially in a short time frame. When we have to eventually recommend tweets, we can perform an inverse mapping from topics to current tweets (appropriately taking into account the many-to-one aspect via sampling or other procedures). Thus, for the majority of the paper, we assume that the users engage with the topics instead of tweets (e.g., via reply, like, retweet, or retweet with comment engagement behaviors) at every time period (i.e., week, day, hour). We use a binary vector $\textbf{e}_{it} = [ e_{it0},....,e_{itj}] \in \left\{0,1\right\}^{J \times 1} $ to denote the engagement of user $i$ at time $t$. The binary indicator $e_{itj} \in \left\{0,1\right\}$ represents whether user $i$ engaged with topic $j$ at time $t$.

We summarize information about past engagement behavior of user {i} by an immediate engagement history of length $T$, as well as by using the topic engagement frequencies over the entire available time horizon. We denote the engagement history of length $T$ for user $i$ at time  $t$ by $E_{it}^{T} = [\textbf{e}_{i,t}, \textbf{e}_{i,t-1},...,\textbf{e}_{i,t-T+1}] \in \left\{0,1\right\}^{J \times T}$ and the vector of topic-specific engagement frequencies for user $i$ over the entire user engagement history available at time $t$ by $E_{it}^{\infty} = [ \bar{e}_{it0},....,\bar{e}_{itj}] \in \left[0,1\right]^{J \times 1} $. Users receive topic recommendations during each time period (as mentioned before, these will be mapped to available tweets on their home timeline screen/tab). We denote these recommendations by ${R}_{it} = [ r_{it0},....,r_{itj}] \in \left\{0,1\right\}^{J \times 1} $, where $r_{itj} \in \left\{0,1\right\} $ indicates whether a recommendation was made to the user $i$ in time $t$ for topic $j$.

We propose a choice-aware engagement model that predicts probabilities $P_{i,t+1} = [p_{i,t+1,1},....,p_{i,t+1,j},...,$ $p_{i,t+1,J}]$ with which user $i$ will engage with topics $j \in \left\{1,...,J\right\}$ at time $t+1$, given the current topic recommendations $R_{i,t+1}$, the recent engagement history $E_{it}^{T}$, the longer term engagement frequencies $E_{it}^{\infty}$, and the model parameters $\theta$:
\begin{equation*}
   P_{i,t+1} = f(R_{i,t+1},E_{it}^{T},E_{it}^{\infty},\theta), 
\end{equation*}
where $f$ denotes the choice-aware engagement function parameterized by $\theta$. The vector $P_{i,t+1}$ contains the probabilities for the (binary) engagements for all topics $j$:
\begin{equation*}
   p_{i,t+1,j} = P(e_{i,t+1,j} = 1). 
\end{equation*}
Including both $E_{it}^{\infty}$ and $E_{it}^{T}$ as an input to the model serves two purposes. First the model can use $E_{it}^{\infty}$ to learn base/static preferences of the user, and  it can model engagement patterns that change over time using $E_{it}^{T}$. Separating the information already at the model input simplifies the learning process, affords higher interpretability, and speeds up the training. Second, providing $E_{it}^{\infty}$ in addition to $E_{it}^{T}$ reduces the dimensionality of the input data. Our model would pick up the $E_{it}^{\infty}$ feature/signal directly from $E_{it}^{T}$ if the window length were set to infinity (i.e $T = \infty$). However, recent engagements are typically more relevant to engagement, so we reduce dimensionality by considering a smaller window $T$ and separately including $E_{it}^{\infty}$ as a summary of all older engagements.

The model is defined as a neural network. Each observation used to train our model is a user-time pair $(i,t)$. For every training example, 
the model transforms the inputs (i.e., $R_{i,t+1}, E_{it}^{T}, E_{it}^{\infty}$) to create topic-specific feature maps $\textbf{z}_{i,t+1,j} \in R^{K \times 1} $, which are then used to predict the engagement probabilities $p_{i,t+1,j}$ for every topic in the tweet stream (here $K$ is a parameter of the network). In other words,

\begin{equation*}
    \begin{array}{l}
    p_{i,t+1,j} = p(z_{i,t+1,j};\theta_{p}), \text{with}\\
    \\ \textbf{z}_{i,t+1} = [z_{i,t+1,1},....,z_{i,t+1,J}] \in {\mathbb R}^{J \times K} = f'(R_{i,t+1},E_{it}^{T},E_{it}^{\infty};\theta_{z}), 
    \end{array}
\end{equation*}
where $f'(\cdot;\theta_z)$ above is an intermediate function, which when composed with $p(\cdot;\theta_p)$ results in $f$. The feature maps $z_{i,t+1,j}$ summarize information about the recommendations and the engagement behaviour into user-topic-specific $K$-dimensional vectors. Next, we describe the details of our model architecture and how we train it. In particular, for training, our aim is to  to find the model parameters such that the distribution and behavior of the predicted engagement probabilities is similar to the distribution and behavior of engagement probabilities observed in training data. 
 
\subsection{Model Architecture}

Recall that the inputs to the model are the user topic recommendations $R_{i,t+1}$, the recent engagement history $E_{it}^{T}$ and the longer term topic engagement frequencies $E_{it}^{\infty}$.

Engagement histories $E_{it}^{T}$ are typically sparse. The model first transforms the engagement histories $E_{it}^{T}$ in the following way. We apply convolution operations with $H$ different real-valued filters $\textbf{w}_{h} \in {\mathbb R}^{T \times 1}$ and a leaky $\textbf{ReLU}$ activation function to get denser embeddings:

\begin{equation*}
    E_{it}^{H} = [\sigma(E_{it}^{T} \cdot \textbf{w}_1),....,\sigma(E_{it}^{T} \cdot \textbf{w}_H)] \in {\mathbb R}^{J \times H},
\end{equation*}
where $\sigma(\cdot)$ is the leaky $\textbf{ReLU}$ activation function~ \cite{xu2015empirical}.

The filters apply the same transformations to the history vector corresponding to each topic, and create $H$-dimensional topic-specific summary statistics (or embeddings). These summary statistics represent information about recent engagements in a dense form. We learn the weights of these filters using the training data.

The above approach for summarizing temporal information is more flexible than explicitly pre-defined transformations of the engagement histories (e.g., weighted averages) that have been considered in many previous works. This flexibility is important because tweet-topics may have a substantial variation in terms of their engagement levels across time. For example, users typically engage with topics of greater relevance more frequently and vice-versa. Explicitly pre-defining filter weights to capture such patterns will be challenging. And we avoid this by learning their weights in a data-driven way using engagement patterns in the training data.

Engagement frequencies $E_{it}^{\infty}$ , the transformed engagement histories $E_{it}^{H}$ and topic recommendations $R_{i,t+1}$ are topic-specific. So, we use bottleneck layers in our neural network architecture to share information across topics. In particular, we apply the following transformations:

\begin{equation*}
    \begin{array}{l}
    F_{i,t}^{\infty} = W_{\infty} E_{i,t}^{\infty},\quad  
    F_{i,t}^{H} = W_{H} E_{i,t}^{H},\quad F_{i,t+1}^{D} = W_{d} R_{i,t+1}, \\\\
    
    \bar{E}_{i,t}^{\infty} = W_{\infty}^{'} F_{i,t}^{\infty},\quad  
    \bar{E}_{i,t}^{H} = W_{H}^{'} F_{i,t}^{H}, \textrm{ and} ~ \bar{R}_{i,t+1} = W_{d}^{'} F_{i,t+1}^{D},
    \end{array}
\end{equation*}

where $W_d, W_\infty$ and $W_H$ are $(L \times J)$ weight matrices with $L<<J$, and $W'$ denotes the transpose of matrix $W$. The bottleneck layers encode the inputs into low-dimensional feature rich representations  $F_{i,t}^{\infty}, F_{i,t}^{H} $ and $F_{i,t+1}^{D}$. The model infers the weight matrices $W_{d}, W_{\infty}$, and $W_{H}$ during training.

\begin{figure}
  \includegraphics[width=\linewidth]{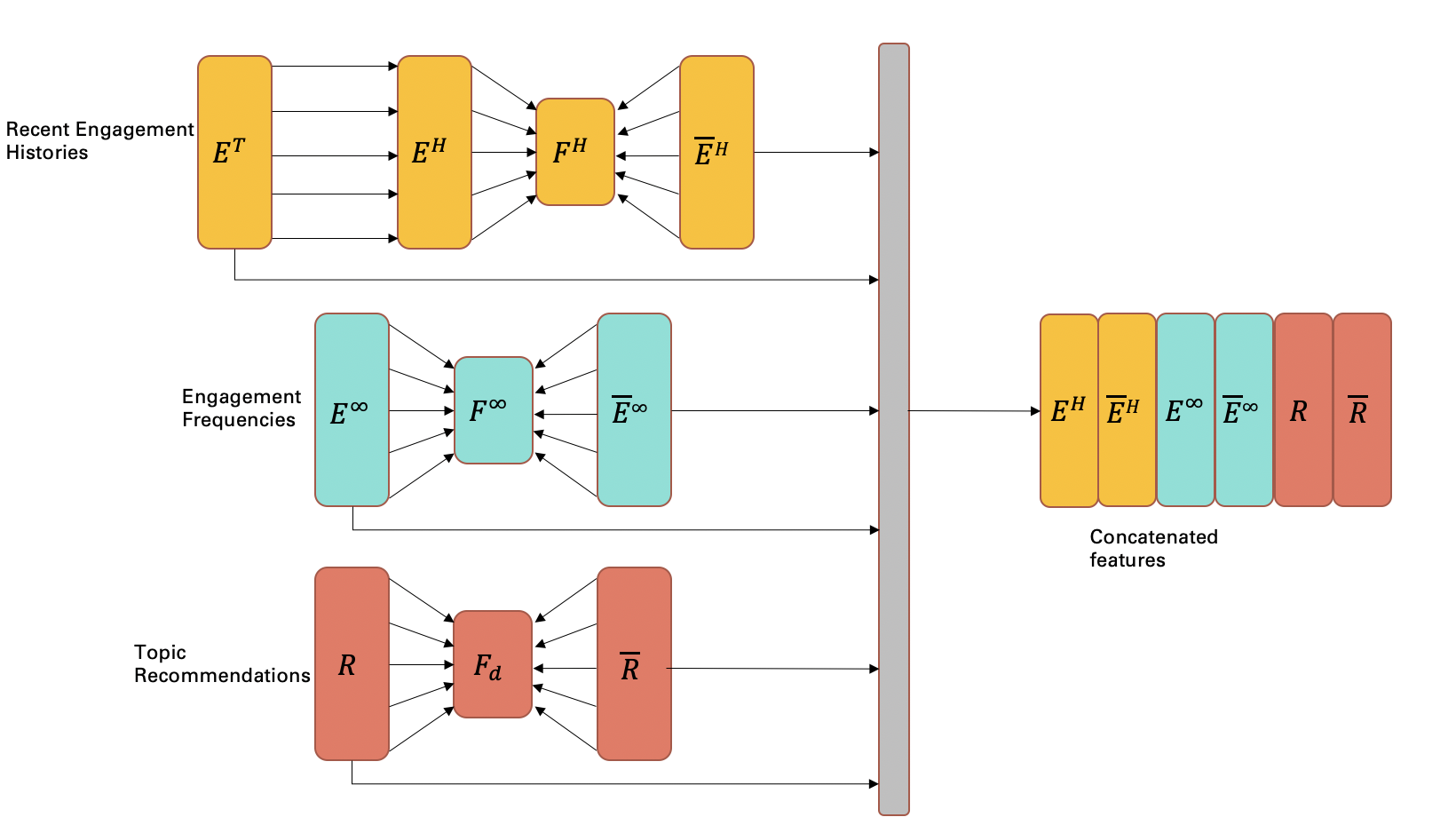}
  \caption{Proposed architecture that captures choice-aware user engagement behavior.}
  \label{fig:model}
\end{figure}

The bottleneck layers described above are the basis for modelling cross-topic relationships. Consider the following illustrative example. User $i$ is indifferent to two tweet-topics $p_1$ and $p_2$. When he/she engages with topic $p_1$ or $p_2$ at time $t$, the model needs to adjust the estimates of the probabilities that the user will engage with these topics at time $t+1$. The adjustment in probabilities should ideally be independent of which particular topic was engaged with in time $t$. The model recognizes this by creating similar $L$-dimensional representations of the engagement histories $E_{it}^{H}$ and the engagement frequencies $E_{it}^{\infty}$ for the two different scenarios (engaging with $p_1$ or $p_2$). These $L$-dimensional representations are then expanded back to $J$ dimensions to continue further modeling at the topic level. See Figure~\ref{fig:model} for an illustration of the bottleneck modules and the concatenated features that are passed through a straightforward softmax layer to obtain engagement probabilities as described next.

In particular, we combine the inputs and outputs of the bottleneck layers (thus using the skip/residual-connection idea from the deep learning literature) to create feature map $\textbf{z}_{i,t+1}$:
\begin{equation}
\label{combined}
    \textbf{z}_{i,t+1} = [\textbf{1}^{J \times 1}, R_{i,t+1}, \bar{R}_{i,t+1}, E_{it}^{\infty}, \bar{E}_{it}^{\infty}, E_{it}^{H} \bar{E}_{it}^{H}] \in {\mathbb R}^{J \times K}. 
\end{equation}

Combining the inputs and outputs of such layers is a standard trick to improve the predictive performance of the neural network. Further, in Equation \eqref{combined}, we capture biases by appending $\textbf{1}^{J \times 1}$ to the combined features. Biases allow for a richer representation of the input, and also tend to improve the overall learning of  the model parameters. The feature maps $\textbf{z}_{i,t+1,j}$ summarize relevant information about the user engagement behavior and the topic recommendations, and a subsequent sigmoid transformation/linear layer uses $\textbf{z}_{i,t+1,j}$ to predict the engagement probability of user $\textbf{i}$ engaging with each topic $\textbf{j}$ at time $\textbf{t}$. The parameters $\theta_p$ are shared between the topics.

\subsection{Model Training}

The parameters of the model are the time filter parameters $\textbf{w}_h$, bottleneck layer parameters $W_d, W_{\infty}$, and $W_H$, and the parameters of the softmax layer $\theta_{p}$:
\begin{equation*}
\theta = (\theta_{z}; \theta_{p}), \textrm{ and } \theta_{z} = (\textbf{w}_{h = 1...H};W_d;W_\infty;W_H).
\end{equation*}
We learn the parameters by minimizing the sum of  binary cross-entropy losses, one per topic. Thus,
\begin{equation}
\label{eqn:bce}
    \begin{array}{l}
    \theta^{*} = \arg\min_\theta \sum_{i=1}^{I}\sum_{j=1}^{J}\sum_{t=1}^{T} L(e_{i,t+1,j}, \hat{p}_{i,t+1,j}), \;\;\text{with}\\
    L(e_{i,t+1,j}, \hat{p}_{i,t+1,j}) = -((e_{i,t+1,j})~ \text{log}~ \hat{p}_{i,t+1,j} + (1-e_{i,t+1,j}) ~\text{log}~(1-\hat{p}_{i,t+1,j})).
    \end{array}
\end{equation}

We use the popular adaptive moment estimation algorithm (Adam)\cite{kingma2014adam} with mini-batches to optimize the parameters. Adam is a gradient descent method that is computationally efficient and computes adaptive learning rates for each parameter of the model to improve learning stability and speed.

The proposed neural network architecture incorporates two natural constraints on the parameters to facilitate faster model convergence and prevent over-fitting. Firstly, recall from earlier, that we assume the weights at the bottleneck layer decoder to be the transpose of the encoder parameters. In particular, we estimate $\bar{R}_{i,t+1} = W_{d}^{'} F_{i,t+1}^{D} = W_{d}^{'} W_{d}{R}_{i,t+1} $ ,where $W_{d}^{'}$ is a transpose of the weight matrix $W_{d}$. The tied weights constraint helps to reduce the number of model parameters and serves as a regularization mechanism. Similarly, we assume tied weights $\theta_{p}$, based on our experimentation. In other words, sigmoid layer applies to the topic-specific activation vectors $\textbf{z}_{i,t+1,j}$, but the parameters $\theta_p$ are shared across all the $J$ topics. 

Overall, our proposed model provides a flexible and partially interpretable functional form that attempts to  closely approximate user engagement behavior. It incorporates information about the engagement history and the topic recommendations to make user-specific and time-specific predictions for every topic. The standalone parameters of the model have no intuitive behavioral or economic interpretations that are readily available, but as seen in our experiments, the model is able to effectively predict engagement behavior taking choice into account, leading to improved personalized tweet recommendations. 

The choice of our neural network architecture also makes the model training computationally tractable and scalable. We train our model in mini-batches to take advantage of stochastic optimization and not have all training data in memory. The proposed neural network architecture allows us to compute gradients via back-propagation efficiently. Training our neural network is therefore feasible even with a large number of users and topic alternatives.

\section{Optimizing Tweets for Engagement Maximization}
\label{optimization}
Generating personalized tweets for a user depends not only on the choice model but also on the downstream tweet assortment computation procedure, which is a combinatorial problem. In particular, the objective is to design an algorithm that allocates a subset of tweet-topics to every user, based on their estimated cross-effects and the overall effect on engagement probabilities. In our approach, the optimization algorithm is kept agnostic to the underlying engagement prediction model, and one can easily swap out existing tweet relevance prediction models with our choice-aware engagement prediction model.

For every user, we aim to select a set of topics that yield the highest engagement uplift. Formally, we solve the following maximization problem:

\begin{equation}
\label{optimequation}
    R_{it}^{*} = \arg\max_{R =[r_1,...,r_J]}\sum_{j} \hat{p}_{itj}(R,E_{it}^{T},E_{it}^{\infty})\times r_{itj},  
\end{equation}
\begin{equation*}
\textrm{ such that} \;\; r_{itj} \in \{0,1\}\; \forall\; (i,t,j),  \;\;\textrm{and} \;\; \sum_j  r_{itj}= n_{it}\; \forall\; (i,t).
\end{equation*}

Here $\hat{p}_{itj}$ is the predicted engagement probability for user $i$ engaging with topic $j$ at time $t$, $r_{itj}$ is the indicator variable that determined if the topic recommendation $j$ was made to user $i$ at time $t$, and $n_{it}$ is the number of topic recommendations that can be made to the user $i$ at time $t$ (this is typically restricted by the device display size of the phone/computer used by the user). Notice that the decision variable $R=[r_1,...,r_J]$ also enters the engagement model, thus making the objective a highly complex nonlinear function of the binary variables. 

Finding the subset of optimal tweet-topics for the problem in Equation~\eqref{optimequation} by exhaustive search is not feasible, since evaluating all the topic combinations for the users needs exponential time. Further, the optimization problem is NP-hard without further assumptions on the engagement model. For instance, if $\hat{p}_{itj}$ was computed based on a simple multinomial logit model\cite{tulabandhula2020optimizing}, then the problem in Equation~\eqref{optimequation} is known to be polynomial time (with exact solutions achievable using custom designed algorithms in time $O(J^2)$).

Thus, we consider a greedy heuristic for topic allocation when using our engagement model. The greedy heuristic begins by selecting a single topic that maximizes engagement likelihood/uplift. It then sequentially adds topics, one topic at a time, to maximize the uplift, i.e., the objective in Equation~\eqref{optimequation}, given the previously chosen topics. The algorithm terminates once the predefined number of topics $n_{it}$ is selected. While the approach is quite straightforward with iteration complexity equal to $n_{it}$, it has been previously shown\cite{berbeglia2020assortment,tulabandhula2020multi} in the retail contexts that such an approach can yield good solutions for many popular choice models.

\section{Experimental Results}
\label{experiments}
To evaluate our proposed engagement model and optimization scheme, we conduct extensive experiments and compare the performance of our model with the baselines. We chose retweet engagement for our analysis in this section, and the same can be easily extended to other forms of engagement. In the following, we describe the dataset used to train and test the model (Section \ref{dataset}), evaluation protocols (Section \ref{evaluation}),  enumerate the parameter settings (Section \ref{parameters}) and baseline methods (Section \ref{baselines}), and finally present our findings (Section \ref{results}).

\subsection{Dataset}
\label{dataset}
We experiment on a large scale dataset made publicly available by Twitter as part of the RecSys Challenge 2020 \cite{belli2020privacyaware}. The dataset has 160 million tweets curated along with their engagement features from users' home timeline. The dataset is very large and highly sparse. To ensure the quality of the dataset, we perform modest filtering on the data, retaining only English language users with at least 20 tweets on their home timeline. We retrieve the raw tweet text by de-tokenizing the BERT ids \cite{devlin2018bert} from the filtered dataset. We perform standard data pre-processing, including stop word removal and stemming on the raw text. We cluster the tweets by topics, and label them using the GSDMM \cite{yin2014dirichlet} topic modeling framework. The allocation of tweets by the topic label for 10 of the most prominent topics are shown in Table \ref{tab:table1}. We omit the interpretation of the resulting clusters here, and refer the reader to \cite{yin2014dirichlet} for more details. The resulting pre-processed dataset contains 107,017 users and 2.7 million tweet samples. The following features characterize each data entry:
\begin{itemize}[noitemsep]
\item User-id,
\item Tweet-id,
\item Tweet-label,
\item Tweet timestamp, and
\item Retweet timestamp.
\end{itemize}

We split the resulting dataset into 14 time periods of 12 hours each. We define a quantity called the \emph{active state}, which is  the period between when a tweet is published and when it is engaged with. For every user, we consider positive examples from all the tweets posted during their active states and vice versa.  We consider recent engagement history from 4 time periods and engagement frequency from all the time-periods available up to now. For training, validation, and testing, we compare our predictions with the ground truth label (the observed engagement) from the respective time periods. 
\begin{table}
\caption{Top 10 tweet-topic labels by number of tweets assigned to a topic.}
\label{tab:table1}
\centering
\begin{tabular}{cclllllllllll}

\cline{1-2}
\multicolumn{1}{|l|}{Tweet-topic Label} & \multicolumn{1}{c|}{Number of Tweets} &  &  &  &  &  &  &  &  &  &  &  \\ \cline{1-2}
\multicolumn{1}{|c|}{29}           & \multicolumn{1}{c|}{547139}       &  &  &  &  &  &  &  &  &  &  &  \\ \cline{1-2}
\multicolumn{1}{|c|}{9}            & \multicolumn{1}{c|}{481994}       &  &  &  &  &  &  &  &  &  &  &  \\ \cline{1-2}
\multicolumn{1}{|c|}{34}           & \multicolumn{1}{c|}{232720}       &  &  &  &  &  &  &  &  &  &  &  \\ \cline{1-2}
\multicolumn{1}{|c|}{4}            & \multicolumn{1}{c|}{201084}       &  &  &  &  &  &  &  &  &  &  &  \\ \cline{1-2}
\multicolumn{1}{|c|}{7}            & \multicolumn{1}{c|}{147789}       &  &  &  &  &  &  &  &  &  &  &  \\ \cline{1-2}
\multicolumn{1}{|c|}{32}           & \multicolumn{1}{c|}{142026}       &  &  &  &  &  &  &  &  &  &  &  \\ \cline{1-2}
\multicolumn{1}{|c|}{14}           & \multicolumn{1}{c|}{128295}       &  &  &  &  &  &  &  &  &  &  &  \\ \cline{1-2}
\multicolumn{1}{|c|}{8}            & \multicolumn{1}{c|}{126075}       &  &  &  &  &  &  &  &  &  &  &  \\ \cline{1-2}
\multicolumn{1}{|c|}{17}           & \multicolumn{1}{c|}{122814}       &  &  &  &  &  &  &  &  &  &  &  \\ \cline{1-2}
\multicolumn{1}{|c|}{21}           & \multicolumn{1}{c|}{113852}       &  &  &  &  &  &  &  &  &  &  &  \\ \cline{1-2}
\multicolumn{1}{l}{}               & \multicolumn{1}{l}{}              &  &  &  &  &  &  &  &  &  &  &  
\end{tabular}
\vspace{-10mm}
\end{table}

\subsection{Evaluation Protocols}
\label{evaluation}
To evaluate our model, we use two metrics : AUC (Area Under the ROC curve) and  BCE  (Binary Cross-Entropy) loss on the held out test set. These two metrics assess the performance from two different perspectives: AUC measures the probability that a positive instance (i.e., a positively engaged tweet-topic) will be ranked higher than a randomly chosen negative one. A high performing model has AUC closer to 1, which means that it has a good measure of separability. A poor model has AUC closer to 0, which means it is not able to separate positive and negative instances. The main drawback of this measure is that it only considers the order of predicted probabilities and is insensitive to class imbalance. To fix this, one could look at sensitivity at a specific specificity value, but due to the other drawbacks of this latter approach, we choose to report AUC. In contrast, the BCE loss measures the distance between the predicted score and the true label for each instance, and is essentially a negative log likelihood score for the model parameters chosen.

\subsection{Parameter Settings}
\label{parameters}

We implement our neural network model using Pytorch v1.8.1. To achieve our model's best performance, we tune hyper-parameters by optimizing binary cross-entropy loss in Equation \eqref{eqn:bce}. To select the ideal number of convolution filters $H$ , we tested the values from the set $\{5,10,15,20,30\}$ and set it to $20$. We search for optimal batch size and learning rate in sets $\{16, 32, 64, 128\}$ and $\{1e^{-3}, 1e^{-4}, 5e^{-5}, 1e^{-5}, 1e^{-6}\}$, respectively. The learning rate is finally set to 1$e^{-5}$. We use the Adam optimizer with a mini-batch size of $32$. We ran all our experiments for $50$ epochs during training. The experiments are performed on a Linux machine with a $12$ core $32$GB 64-bit Intel(R) Core(TM) i7-8700 CPU @ 3.20GHz processor, and an Nvidia GeForce GTX 1080 GPU. The source code to replicate our experimental results is available at \url{https://github.com/me10b031/twitter_user_model}.

\subsection{Baselines}
\label{baselines}
To compare and contrast the effectiveness of our proposed model, we study the performance of two baselines discussed below.

The first baseline is a binary logit model. We apply the binary logit model per topic independently. For each topic, the independent variables are the user-specific topic engagement frequencies $\bar{e}_{itj}$ , the engagement histories $[\textbf{e}_{i,t}, \textbf{e}_{i,t-1},..,$ $\textbf{e}_{i,t-T+1}]$ and the topic recommendations $r_{i,t+1,j}$.

We use a gradient-boosted decision tree classifier (implemented using LightGBM) as a second baseline. LightGBM is an efficient implementation of the gradient boosting decision tree algorithm. We estimate the model parameters with the same training dataset as the one used for the binary logit model.

\subsection{Findings}
\label{results}

Table \ref{tab:table2} illustrates the prediction performance of our proposed neural network model. We report the average binary cross-entropy loss and average AUC scores calculated using predicted engagement probabilities on test data. Our model achieves better average predictive performance than the baseline models on both metrics (lower in terms of BCE loss and higher in terms of AUC). 

\begin{table}
\caption{Performance metrics across the proposed and the baseline models.}
\label{tab:table2}
\centering
\begin{tabular}{|l|l|l|}
\hline
              & BCE loss &  AUC    \\ \hline
Our Model     & \textbf{1.095}    & \textbf{0.8972} \\ \hline
Binary Logit & 1.146    & 0.8061 \\ \hline
LightGBM      & 1.188    & 0.5451 \\ \hline
\end{tabular}
\vspace{-6mm}
\end{table}

One reason why our neural network model performs better is the following: our model extends the baselines by using information from all topics simultaneously as input to predict engagement for each topic. Leveraging rich high-dimensional information for all topics is possible with the proposed model architecture. It includes data driven time filters to reduce engagement history sparsity, bottleneck layers to encode cross-topic relationships, and weight sharing to minimize the parameters and regularize the model. Most importantly, it includes the impact of other topics through vector $R$, and thus captures substitution/complementarity effects that ultimately influence engagement with each chosen topic.

We show average engagement uplift scores (the objective function in Equation~\eqref{optimequation}) considering five tweet-topic recommendations (i.e., $n_{it} = 5$) per user in Table \ref{tab:table3}. As expected, our model yields higher engagement uplift scores compared to the recommended tweet-topics produced using the baseline models. In this table, each model produces a recommended set of five tweet-topics using the greedy approach outlined in Section~\ref{optimization}. These recommendations are then evaluated against the objective function defined using the most complex model (namely the proposed model). We do note that this approach has a bias towards our model. Nonetheless, it can still be considered as an approximate/proxy measure for evaluating the aggregate improvement in engagement. While it may seem from the table that the scores are not very different between the three models, these scores are the sum of the probabilities of engaging with each of the tweet-topics and can only vary between $[0,5]$. As such, even a minor improvement in this score (as little as $2\%$ relative value as can be deduced from the table) can have an out-sized impact on the platform's bottom-line at scale.

\begin{table}
\caption{Engagement uplift scores for optimized tweet-topic recommendations across models.}
\label{tab:table3}
\centering
\begin{tabular}{|l|c|}
\hline
              & \begin{tabular}[c]{@{}c@{}}Engagement Uplift \\ Score\end{tabular} \\ \hline
Our Model     & \textbf{1.677}                                                              \\ \hline
Binary Logit & 1.644                                                             \\ \hline
LightGBM          & 1.636                                                             \\ \hline
\end{tabular}
\vspace{-6mm}
\end{table}

\section{Conclusion}
\label{conclusion}
In this paper, we have designed a neural network model to predict tweet choice behavior given a collection of tweets. The model is motivated by the problem of showing a personalized timeline of tweets to each user on the Twitter platform t any given point of time. Given candidate tweet-topic recommendations and user engagement histories, our model predicts individual choice-aware engagement probabilities for every tweet-topic in the tweet stream applicable to this user. Our model eliminates the need for pre-defined topic attributes and instead learns them in an unsupervised manner. It also captures cross-topic effects inherently due to the design of our neural network architecture.

We have evaluated the predictive performance of our model in our experiments with a large scale Twitter dataset. The model outperforms standard machine learning benchmarks both in prediction as well as when used for optimizing topic recommendations. This improved performance can be attributed to the model being able to capture approximate cross-topic effects out-of-sample. 

Overall, explicit accounting of choice behavior (i.e., the influence of a user engaging with a tweet depends on other tweets being shown at the same time) is significant given limited screen-sizes, and limited user attention. It can have an out-sized impact on business goals, and as we show here, can be captured in an algorithmic decision-making process. While traditional choice models in the marketing literature have been somewhat simplistic in nature (such as the multinomial logit, or the distribution over rankings choice model for instance), our use of neural networks to capture choice, especially in a recommendation focused application, is relatively novel. We can apply our model design, data treatment, and optimization approaches to other user engagement problems faced by online Internet services in a similar manner.

\bibliographystyle{splncs04}
\bibliography{references}

\end{document}